%% file: new_main.tex
\documentclass[a4paper]{article}

\usepackage{INTERSPEECH2021}
\usepackage{multirow}
\usepackage[a-1b]{pdfx}   %

\usepackage[utf8]{inputenc}
\usepackage{kotex}
\usepackage{kotex-logo}

\usepackage[noadjust]{cite} %

\usepackage{algorithm,algpseudocode}%

\title{GANSpeech: Adversarial Training for High-Fidelity \\  Multi-Speaker Speech Synthesis}
\name{Jinhyeok Yang$^*$, Jae-Sung Bae$^*$, Taejun Bak, Youngik Kim, Hoon-Young Cho\thanks{$^*$ Equal contribution.}}

\address{
    Speech AI Lab, NCSOFT, Republic of Korea
}
\email{\{yangyangii,jaesungbae,happyjun,youngik,hycho\}@ncsoft.com}

\begin{document}

\maketitle
\begin{abstract}
Recent advances in neural multi-speaker text-to-speech (TTS) models have enabled the generation of reasonably good speech quality with a single model and made it possible to synthesize the speech of a speaker with limited training data. Fine-tuning to the target speaker data with the multi-speaker model can achieve better quality, however, there still exists a gap compared to the real speech sample and the model depends on the speaker.
In this work, we propose GANSpeech, which is a high-fidelity multi-speaker TTS model that adopts the adversarial training method to a non-autoregressive multi-speaker TTS model. In addition, we propose simple but efficient automatic scaling methods for feature matching loss used in adversarial training. 
In the subjective listening tests, GANSpeech significantly outperformed the baseline multi-speaker FastSpeech and FastSpeech2 models, and showed a better MOS score than the speaker-specific fine-tuned FastSpeech2.
\end{abstract}

\section{Introduction}
\label{sec:intro}
Recent advances in deep learning enabled the neural text-to-speech (TTS) systems to synthesize realistic and natural speech. The neural TTS systems typically first generate a mel spectrogram from given text \cite{wang2017tacotron, shen2018tacotron2, deepvoice2, deepvoice3, ren2019fastspeech, ren2021fastspeech2, fastpitch, jsbae} first, and consequently synthesize a speech waveform from the mel spectrogram using a vocoder \cite{oord2016wavenet, oord2018parawavenet, yang2020vocgan, yamamoto2020parallel, kumar2019melgan, hifigan}. In the rest of this paper, we refer to TTS as the mel spectrogram generation part. Nowadays, the non-autoregressive (non-AR) neural TTS \cite{ren2019fastspeech, ren2021fastspeech2, fastpitch} generates comparable speech quality with lesser pronunciation error and faster synthesizing speed than the previous autoregressive (AR) neural TTS \cite{wang2017tacotron, shen2018tacotron2, deepvoice2, deepvoice3}. With enough recorded speech data, nowadays, the neural TTS model can generate natural sounding, human-like speech.

The extension of the TTS model to make it capable of synthesizing speech of multiple speaker voices in a single model has been widely studied in both autoregressive \cite{deepvoice2, deepvoice3, libri_tts, multispeech, towards_univ_tts} and non-autoregressive TTS models \cite{fastpitch, adaspeech}. These multi-speaker TTS (MS-TTS) models are more efficient than building single speaker models for each speaker and make it possible to synthesize the speech of a speaker with only a small amount of speech data (which is impossible in a single-speaker model \cite{semi_sup_improving_data_eff}). 
More recently, few-shot \cite{chen2018sample_efficient, neural_voice_cloning, high_quality_lightweight, boffintts, zhang2020adadurian, adaspeech} and zero-shot \cite{transfer_learning_from_ASV, zero_shot_MS} speaker adaptation on MS-TTS model have been studied. These models enable the generation of speech of a speaker that only has a few minutes of speech data.

To obtain high performance in few-shot and zero-shot speaker adaptation, the well pre-trained MS-TTS model is essential. To build a high-fidelity MS-TTS model, some approaches adopted an externally pre-trained automatic speaker verification (ASV) model \cite{transfer_learning_from_ASV, zero_shot_MS}. However, they need additional domain knowledge and require external speech data to train an ASV model. These studies are effective for unseen speakers, but the performance for seen speakers is hardly improved compared to a method using an embedding look-up table with speaker labels \cite{deepvoice2, deepvoice3, multispeech}.
In previous studies \cite{high_quality_lightweight,song2020speakeradaptive}, the authors further improved the performance of a target speaker voice by fine-tuning the pre-trained multi-speaker model to the limited target speaker-specific dataset. However, there still exists a gap compared to the real speech sample and the fine-tuned model depends on the target speaker.

The generative adversarial network (GAN) showed powerful performance in several image domains \cite{wang2018pix2pixhd,zhang2018stackgan++}, and was adopted successfully in speech domain. Especially, the GAN-based neural vocoder models \cite{yang2020vocgan, yamamoto2020parallel, kumar2019melgan, hifigan} showed comparable performance to the AR state-of-the-art vocoder model \cite{oord2016wavenet} and can synthesize speech in real-time. MelGAN \cite{kumar2019melgan} successfully applied the feature-matching loss \cite{salimans2016improved}, which is loss using the intermediate features of discriminator, to the vocoder model. ParallelWaveGAN \cite{yamamoto2020parallel}, VocGAN \cite{yang2020vocgan}, and HiFi-GAN \cite{hifigan} also added additional loss functions to improve the speech quality of the vocoder.

Recently, the research adopting GAN for TTS models have been also studied. In \cite{adv_korean_singing}, the authors applied the adversarial training to an super-resolution network converting mel spectrogram that generated by the AR mel-synthesis network to linear spectrogram. In addition, in \cite{sheng2019enhancer}, the authors found that the mel spectrogram generated by the TTS model trained with only reconstruction loss was over-smooth, and proposed an enhancer model for mel spectrogram that reduces the gap between the true mel spectrogram and the generated mel spectrogram. %

\begin{figure*}[h]
  \centering
  \includegraphics[width=0.87\linewidth]{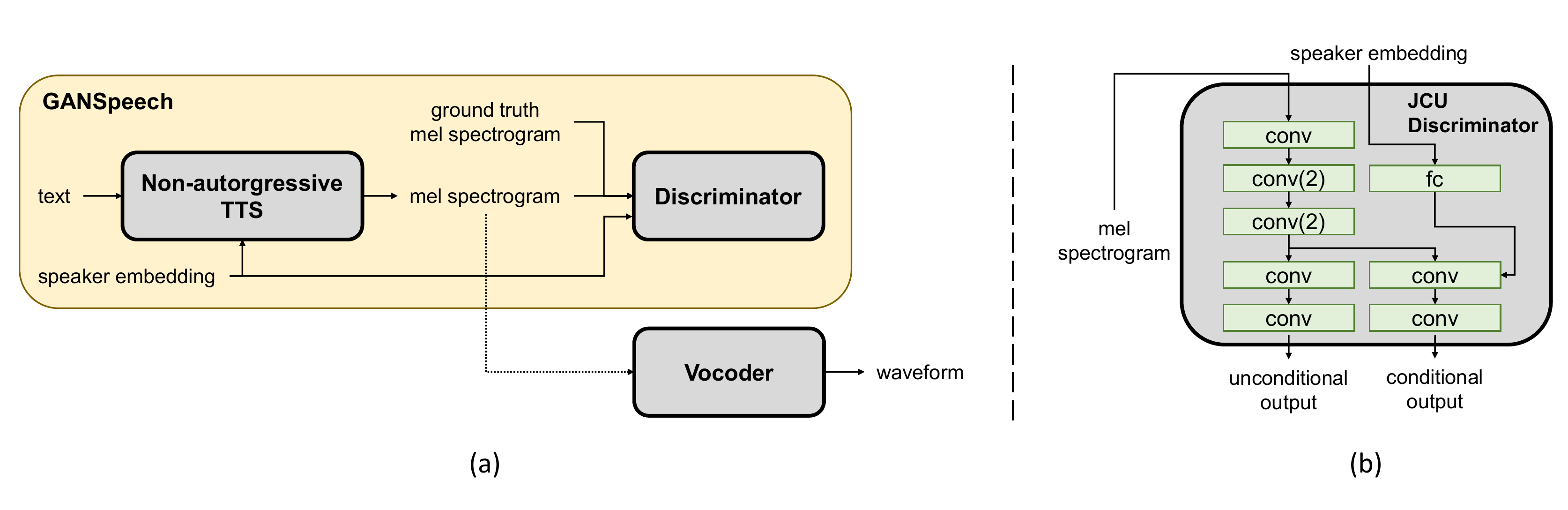}
  \caption{(a) GANSpeech overview. The dotted line is inference only. (b) Discriminator architecture. conv($v$) is convolutional layer with a stride of $v$.}
  \label{fig:1}
\end{figure*}

In this paper, we propose GANSpeech that can synthesize high-fidelity speech of multiple speakers as a single model without an additional module for enhancing TTS model or any speaker-specific fine-tuning. It is trained in two phases. In the first phase, the MS-TTS model is pre-trained using only reconstruction loss. In the second phase, the discriminator and adversarial loss functions are adopted and jointly trained. For the discriminator, we used the joint conditional and unconditional (JCU) discriminator \cite{zhang2018stackgan++} using speaker embedding as a condition, so that it can learn the general characteristic of mel spectrogram along with speaker specific characteristics.
With GANSpeech, we can significantly improve the overall speech quality of the MS-TTS model. For stable GAN training, we also proposed the scaled feature matching loss and reduced the cost of additional hyper-parameter searching for GAN training. We adopted our GANSpeech technique to FastSpeech and FastSpeech2 models and conduct several subjective evaluations to verify the effectiveness of our GANSpeech. The results showed that GANSpeech has better performance than the baseline models and is even better than the fine-tuned FastSpech2 to a specific target-speaker. In addition, we adopted the scaled feature matching loss to VocGAN and showed that it can also be adopted into the GAN-based neural vocoder.

This paper is organized as follows. Section 2 describes the baseline model. In section 3, the proposed GANSpeech model is described. The experimental setup and results are presented in Section 4. Finally, we conclude our work in Section 5.

\section{Baseline Model}
\label{sec:baseline}

For the baseline TTS model, we choose FastSpeech \cite{ren2019fastspeech} and FastSpeech2 \cite{ren2021fastspeech2} which are widely used non-AR TTS model. We use the baseline model as a generator of GAN. FastSpeech is a feed-forward network based on Transformer \cite{transformer}. Both encoder and decoder are composed of 6 Feed-Forward Transformer (FFT) blocks, respectively. Each block consists of self-attention and 1-D convolution. The length regulator that feeds a hidden state from the encoder to the decoder expands the hidden state by the frame length of mel spectrogram corresponding to a character (or phoneme). FastSpeech2 is an improved version of FastSpeech. FastSpeech2, which is a state-of-the-art non-AR model, eases the one-to-many mapping problem by adding a network to predict the variance information such as pitch and energy.

To extend the FastSpeech and FastSpeech2 model to MS-TTS model, we add speaker embedding after the encoder of the model. In \cite{fastpitch} and \cite{zhang2020adadurian}, the speaker embedding was added before and after the encoder, respectively. In our preliminary experiment, we found that adding it after the encoder gave a slightly better performance. In this work, the speaker embedding is obtained by a learnable lookup table.

The baseline models are trained with reconstruction losses. 
The reconstruction loss function of FastSpeech is a summation of mel spectrogram and duration prediction loss. The mel spectrogram prediction loss is the $L^1$ loss between the predicted and the ground truth mel spectrogram and the duration prediction loss is the $L^2$ loss between the predicted and the target duration. FastSpeech2 is trained using additional prediction losses of acoustic features (i.e. pitch and energy) compared to FastSpeech. The additional prediction losses are $L^2$ losses between the predicted and the ground-truth acoustic features. The reconstruction loss function of FastSpeech2 is as follows:
\begin{multline}
    \label{eq:1}
    L_{recon}(G) = L_{mel}(x, \hat{x}) + \lambda_d L_{duration}(d, \hat{d}) \\
                    + \lambda_p L_{pitch}(p, \hat{p}) + \lambda_e L_{energy}(e, \hat{e})
\end{multline}
where $x$, $d$, $p$, and $e$ are target mel spectrogram, duration, pitch, and energy, respectively, and $\hat{x}$, $\hat{d}$, $\hat{p}$, and $\hat{e}$ are corresponding predicted values. 
$\lambda_d$, $\lambda_p$, $\lambda_e$ are the weight of duration, pitch, and energy prediction losses.

\section{GANSpeech}
\label{sec:method}

\input{algo_proposed}

The overall architecture of GANSpeech is depicted in Figure 1, and the training process of GANSpeech is represented in Algorithm \ref{algo_proposed}. It is trained in two phases. First, the pre-training phase that pre-trained the baseline MS-TTS model, generator G. Second, in the adversarial training phase, the model is trained jointly with the reconstruction loss by adding an adversarial loss using a mel spectrogram discriminator D, which is conditioned by a speaker embedding. In the second phase, we also adopted the scaled feature-matching loss that automatically finds an appropriate weight ratio between the feature-matching loss and the reconstruction loss. Two-stage training has the advantages of improving the stability and the training convergence speed, and being able to utilize existing pre-trained TTS models.

\subsection{Discriminator}
\label{subsec:jcu}
We improve the model by adding only the discriminator and applying the adversarial training while using the TTS model (i.e., generator in the adversarial network) of the baseline method as it is. JCU discriminator and loss showed high performance with hierarchically-nested structure\cite{yang2020vocgan,zhang2018stackgan++}. We use the JCU discriminator with a similar structure as VocGAN \cite{yang2020vocgan}. Unlike VocGAN, however, we do not use the hierarchically-nested structure or the multi-scale structure. As shown in Figure \ref{fig:1}, the discriminator consists 1-D convolutions and leaky ReLU activation functions ($\alpha = 0.2$). For the unconditional part of the discriminator, the channels of 1-D convolutions are 64, 128, 512, 128 and 1. The kernel sizes are 3, 5, 5, 5, and 3 and the strides are 1, 2, 2, 1, and 1. We use the speaker embedding as the condition for conditional loss. For the conditional part, it shares three 1-D convolutions with the unconditional part. The speaker embedding is expanded on the time axis after a fully-connected layer with leaky ReLU, and an unconditional output vector is jointly fed to last two 1-D convolutions. The two convolutions have the same architecture as those of the unconditional parts.

\subsection{Adversarial Loss}
\label{subsec:training}

After the model learns the major information of the mel spectrogram using the reconstruction loss, we jointly use the adversarial loss with the reconstruction loss. We use the least-squares loss \cite{mao2017least} with the JCU loss for adversarial training. In each training iteration, the discriminator is trained with Eq.~(\ref{eq:2}) first, and the generator (i.e. TTS model) is trained with Eq.~(\ref{eq:3}), where $s$ is a speaker embedding.

\begin{multline}
    \setlength\abovedisplayskip{0pt}
    \setlength\belowdisplayskip{0pt}
    \label{eq:2}
    L_{\mathrm{D}}^{JCU}(G,D) = \frac{1}{2}\mathbb{E}_s[D(\hat{x})^2 + D(\hat{x},s)^2]\\
    + \frac{1}{2}\mathbb{E}_{(x,s)}[(D(x)-1)^2+(D(x,s)-1)^2]
\end{multline}
\begin{equation}
    \setlength\abovedisplayskip{0pt}
    \setlength\belowdisplayskip{0pt}
    \label{eq:3}
    L^{JCU}_{\mathrm{G}}(G, D) = \frac{1}{2}\mathbb{E}_s[(D(\hat{x})-1)^2+(D(\hat{x},s)-1)^2]
\end{equation}

\begin{figure}[t]
  \centering
  \centerline{\includegraphics[width=6.8cm]{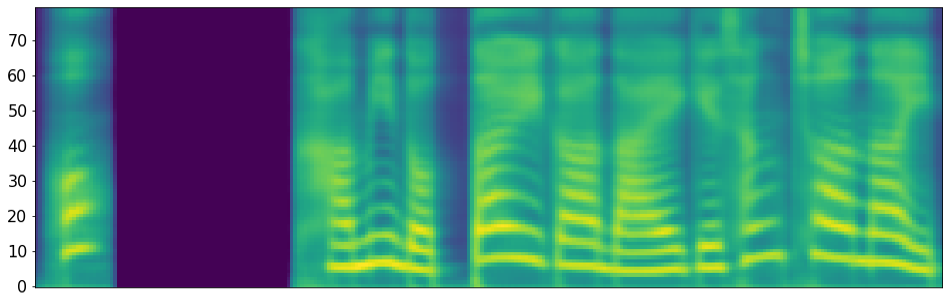}}
  \vspace{-0.05in}
  \centerline{(a)}
  \centerline{\includegraphics[width=6.8cm]{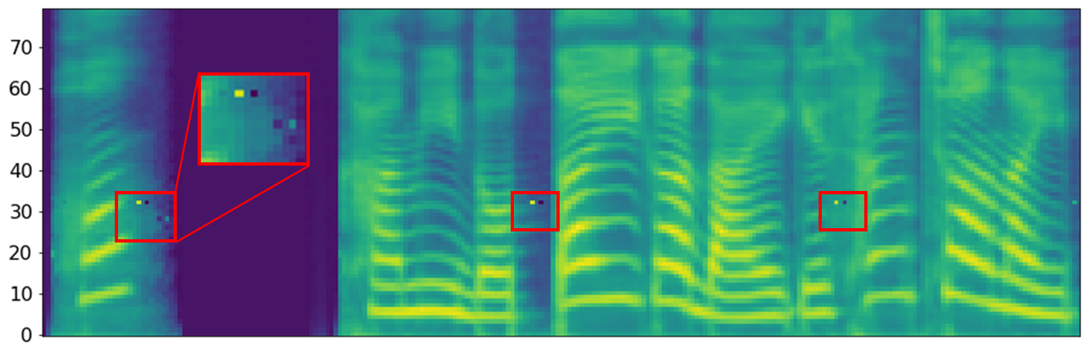}}
  \vspace{-0.05in}
  \centerline{(b)}
  \centerline{\includegraphics[width=6.8cm]{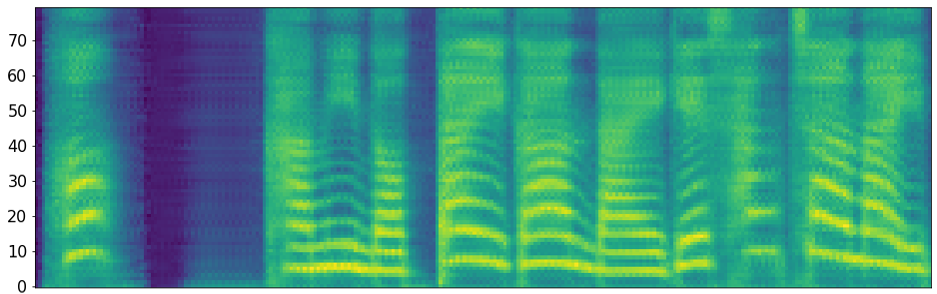}}
  \vspace{-0.05in}
  \centerline{(c)}
  \centerline{\includegraphics[width=6.8cm]{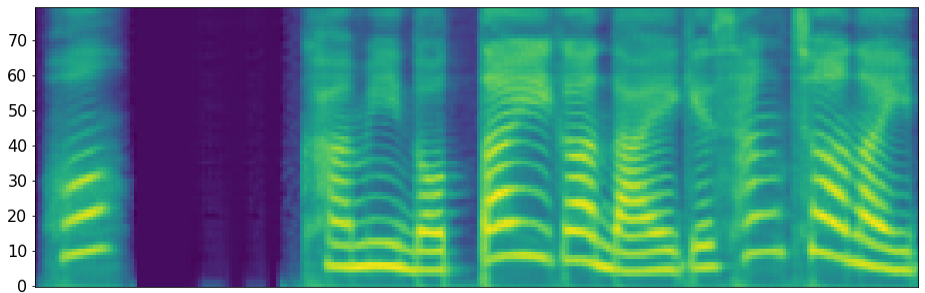}}
  \vspace{-0.05in}
  \centerline{(d)}
\caption{Mel spectrogram generated from the same sentence by each method. (a) FastSpeech. (b) GANSpeech without feature matching loss. (c) GANSpeech with feature matching loss. (d) GANSpeech with scaled feature matching loss}
\label{fig:2}
\end{figure}

\subsection{Scaled Feature Matching Loss}
We also apply the feature matching loss to improve the quality and stability. The feature matching loss successfully improved the quality of GAN-based models not only in the image domain \cite{wang2018pix2pixhd,salimans2016improved} but also in the speech domain \cite{yang2020vocgan,kumar2019melgan}.

\begin{equation}
    \label{eq:4}
    L_{\text{FM}}(G, D) = \mathbb{E}_{(x,s)}\bigg[\sum_{t=1}^{T} \frac{1}{N_t}||D^{(t)}(x) - D^{(t)}(\hat{x})||_1\bigg]
\end{equation}
$T$ is the total number of layers in the discriminator, and $N_t$ is the number of elements in each layer.

This objective has shown good performance and stability empirically, but there is no guarantee that it will reach a fixed point \cite{salimans2016improved}. In addition, because the loss utilizes the intermediate layers of the discriminator, it tends to increase during the training of the discriminator at each step. The gradual increase in the magnitude of the loss can hinder the model from learning key information from the reconstruction loss. In the literatures of \cite{wang2018pix2pixhd,kumar2019melgan}, the ratio of the feature matching loss $\lambda_{\mathrm{FM}}$ is widely used as a fixed hyperparameter 10. However, when additional major losses are jointly used, it is difficult to balance the losses during training. To address this issue, we introduce a scaled feature matching loss that dynamically scales the magnitude of the feature matching loss during training. As shown in line 13 of Algorithm \ref{algo_proposed}, the feature matching loss is adjusted according to $\lambda_{\mathrm{FM}}$, the ratio of the reconstruction loss to the feature matching loss, at each step. The total loss is the summation of the reconstruction loss, the scaled feature matching loss, and the adversarial loss, as in Eq. (\ref{eq:5}).

\begin{equation}
    \label{eq:5}
    L^{total}_G(G, D) = L^{JCU}_{G}(G, D) + \lambda_{\mathrm{FM}} L_{\mathrm{FM}}(G,D) + L_{recon}(G)
\end{equation}

\section{Experiments}

\subsection{Experimental setup}

\subsubsection{Data}

For experiments, we used an studio-quality internal dataset comprising data from  113 Korean speakers, of which 44 were female and 69 were male. The amount of data per speaker varied from 1 minute to 21 hours, totaling 190 hours. The dataset contained 141k scripts and accompanying audio samples. We split 1410 utterances for validation and testing. All remaining samples were used for training. We downsampled the speech data to 22050 Hz for training. We extracted the mel spectrograms from the raw speech waveforms following a previous study \cite{shen2018tacotron2}, and frame length and hop length were set to 1024 and 256, respectively.

\begin{table}[t]
  \caption{Naturalness MOS with 95\% confidence intervals. VocGAN was used as a vocoder of all models. MS: multi-speaker model. FT: speaker-specific fine-tuned model. GT: ground truth. }
  \label{tab:1}
  \centering
  \begin{tabular}{cccc}
    \toprule
    \textbf{Index} & \textbf{Method} & \textbf{MOS} \\
    \midrule
    1   &   FastSpeech (MS)         &   3.10$\pm$0.06   \\
    2   &   FastSpeech2 (MS)        &   3.95$\pm$0.07    \\
    3   &   FastSpeech2 (FT)        &   4.23$\pm$0.06    \\
    4   &   FastSpeech + GANSpeech   &   4.34$\pm$0.06  \\
    5   &   FastSpeech2 + GANSpeech  &   \textbf{4.36$\pm$0.06}   \\
    \hline
        &   GT mel + VocGAN         &   4.50$\pm$0.06    \\
        &   GT wav                  &   4.65$\pm$0.06    \\
    \bottomrule
  \end{tabular}
\end{table}

\subsubsection{Experimental Settings}
All models were trained for 200k steps with a batch size of 32 on an NVIDIA Tesla V100 GPU. We used the Adam optimizer \cite{kingma2014adam} with a learning rate of 0.0001 with $\beta1 = 0.5$ and $\beta2 = 0.9$ for both generator and discriminator. The learning rate was reduced by half every 50k steps. We applied a reduction factor\cite{wang2017tacotron} 2 to all models to reduce the burden on the time length of the mel spectrogram. FastSpeech and FastSpeech2 were learned for all speakers, and FastSpeech2 was fine-tuned for 4 evaluation speakers for optimal performance, respectively. The target durations for training the duration predictor of the baseline models was extracted from attentions of an pre-trained AR Transformer-based TTS model \cite{li2019transformertts}. We used only single encoder-decoder attention, instead of multi-head attention, to get the elaborate duration, and applied location-sensitive attention \cite{shen2018tacotron2, chorowski2015attention} to it.
For training the pitch predictor of FastSpeech2 model, we extracted the pitch information from the raw waveform using Praat \cite{praat}, which is an opensource toolkit for pitch extraction.

We used VocGAN \cite{yang2020vocgan} as a vocoder. Each residual stack of the original VocGAN has 3 dilated convolution blocks. However, we used 7 dilated convolution blocks to improve the speech quality. The other settings are same as the original VocGAN model described in the paper. Additionally, we also applied the proposed scaled feature matching loss to VocGAN. We trained VocGAN for 1M steps with a batch size of 256 on the same internal dataset.

\subsection{Results}

\subsubsection{Mel spectrogram comparison}
Figure \ref{fig:2} illustrated the example mel spectrogram of baseline multi-speaker FastSpeech, GANSpeech without feature matching loss, GANSpeech with feature matching loss without automatic scaling, and GANSpeech with scaled feature matching loss.
As shown in Figure \ref{fig:2}, the mel spectrogram generated by FastSpeech (a) was smoothed and the harmonic structure was not well maintained up to the high frequency. The mel spectrogram synthesized by the GANSpeech (b) was clear and had a harmonic structure up to high frequency range, but had unknown artifacts without the feature matching loss. When we used the feature matching loss (c), the model did not produce the artifacts and improved the performance in the early stage, but it rapidly collapsed during training. Finally, when using the proposed scaled feature matching loss (d), the model converged without disturbing training by the reconstruction loss. The mel spectrogram generated by the proposed method showed more clear and closer to the true mel spectrogram.
\\

\subsubsection{Subjective Evaluation}
We conducted mean opinion score (MOS) test on naturalness for subjective evaluation\footnote{The audio samples are available at the following URL:\\ \url{https://nc-ai.github.io/speech/publications/ganspeech/}}. Multi-speaker FastSpeech and FastSpeech2 model, and FastSpeech2 model with speaker-specific fine-tuning were compared with the GANSpeech model with both FastSpeech and FastSpeech2. Ground-truth (GT) mel spectrogram reconstructed with VocGAN and GT waveform were also compared. For the MOS test, we selected four speakers (two female and two male) which have 1-hour of data, respectively. Five sentences were randomly selected from the test dataset, and its texts were used for synthesizing speech. All the models generated speech with same texts (i.e. 20 speech samples were generated for each model). A total 15 native Koreans participated.

As shown in Table \ref{tab:1}, the proposed method significantly improved the speech quality in both FastSpeech and FastSpeech2. When applied to FastSpeech, it achieved a score of 1.24 higher, and when applied to FastSpeech2, it achieved a score of 0.41 higher. In particular, the proposed method achieved a score of 0.13 higher than the speaker-specific FastSpeech2, fine-tuned with a target speaker data.

\begin{table}[t]
  \caption{Objective measure for VocGAN. SFML: whether the scaled feature matching loss is used. PESQ: higher is better. F0 RMSE: lower is better. GT: Ground-Truth}
  \label{tab:2}
  \centering
  \begin{tabular}{c|cccc}
    \textbf{SFML} & \textbf{PESQ} $\uparrow$ & \textbf{F0 RMSE [Hz]} $\downarrow$ & \textbf{MCD [dB]} $\downarrow$\\
    \hline
    No    &   3.525   &   38.703  &  2.597 \\
    Yes   &   \textbf{3.702}   &   \textbf{34.983}  &  \textbf{2.257}\\
    \hline
    GT &   4.5   &   0.0  &  0.0\\
  \end{tabular}
\end{table}

\subsubsection{Adopting Scaled Feature Matching Loss to VocGAN}
To evaluate the effectiveness of the proposed scaled feature matching loss, we also adopted it to VocGAN and compared the performance. VocGAN uses three kinds of losses for training: multi-resolution STFT loss, adversarial loss, and feature matching loss. Instead of the reconstruction loss of GANSpeech, the multi-resolution STFT loss of VocGAN is used for scaling the feature matching loss.
PESQ \cite{pesq}, F0 Root Mean Square Error (RMSE) and Mel Cepstral Distortion (MCD) were used as objective measures. As shown in Table \ref{tab:2}, when using the scaled feature matching loss, the performance of VocGAN was significantly improved in all measures. In particular, the method made the speech perceptually clear by effectively eliminating the short trembling in the voice.

\section{Conclusion}

This paper proposed the adversarial training method for non-AR TTS model. The method improved the speech quality by applying  to FastSpeech and FastSpeech2. The proposed method showed better multi-speaker TTS performance without speaker-specific fine-tuning. We expect to improve the quality for multi-speaker TTS by applying our method to other non-AR TTS models such as \cite{fastpitch}. The proposed scaled feature matching loss was also shown to be effective when applied to VocGAN. In the future, we expect to improve various GAN models using both the reconstruction loss and the feature matching loss.

\bibliographystyle{IEEEtran}

\bibliography{new_main}

\end{document}

%% file: algo_proposed.tex
\newlength{\continueindent}
\setlength{\continueindent}{2em}
\makeatletter
\newcommand*{\ALG@customparshape}{\parshape 2 \leftmargin \linewidth \dimexpr\ALG@tlm+\continueindent\relax \dimexpr\linewidth+\leftmargin-\ALG@tlm-\continueindent\relax}
\makeatother

\renewcommand{\algorithmicensure}{\textbf{Begin}} %
\newcommand{\Beegin}{\textbf{Beegin}} %
\begin{algorithm}[!t] %
   \caption{Proposed adversarial training algorithm.}\label{algo_proposed} 
  \begin{algorithmic}[1]
  \Statex
    \Require
    \Statex The speech dataset $ S$ composed of text $\mathbf{t}$, mel spectrogram $\mathbf{x}$, speaker $\mathbf{s}$, duration $\mathbf{d}$, pitch $\mathbf{p}$, energy $\mathbf{e}$.
    \State{\textbf{Stage 1: Training only TTS model, generator}}
    \For{number of training epochs}
        \For{$(\mathbf{t}_{i}, \mathbf{x}_{i}, \mathbf{s}_{i}, \mathbf{d}_{i}, \mathbf{p}_{i}, \mathbf{e}_{i}) \in S$}
        \State Calculate $(\mathbf{\hat{x}}_{i},\mathbf{\hat{d}}_{i},\mathbf{\hat{p}}_{i},\mathbf{\hat{e}}_{i})$ from $\textbf{G}$
        \State Update $\textbf{G}$ using $\mathcal{L}_{\mathrm{recon}}$ by Eq.~(\ref{eq:1})
      \EndFor
    \EndFor
    \State{\textbf{Stage 2: Training with adversarial network}}
    \For{number of training epochs}
    
        \For{$(\mathbf{t}_{i}, \mathbf{x}_{i}, \mathbf{s}_{i}, \mathbf{d}_{i}, \mathbf{p}_{i}, \mathbf{e}_{i}) \in S$}
        \State Calculate $(\mathbf{\hat{x}}_{i},\mathbf{\hat{d}}_{i},\mathbf{\hat{p}}_{i},\mathbf{\hat{e}}_{i})$ from $\textbf{G}$
    
        \State Update $\textbf{D}$ using $\mathcal{L}_{\mathrm{D}}$ by Eq.~(\ref{eq:2})
        
        \State $\lambda_{\mathrm{FM}} \gets \mathcal{L}_{\mathrm{recon}} / \mathcal{L}_{\mathrm{FM}}$
        
        \State Update $\textbf{G}$ using $\mathcal{L}_{\mathrm{G}}^{total}$ by Eq.~(\ref{eq:5})
        
      \EndFor
    \EndFor
  \end{algorithmic}
\end{algorithm}

%% file: new_main.bbl
\begin{thebibliography}{10}
\providecommand{\url}[1]{#1}
\csname url@samestyle\endcsname
\providecommand{\newblock}{\relax}
\providecommand{\bibinfo}[2]{#2}
\providecommand{\BIBentrySTDinterwordspacing}{\spaceskip=0pt\relax}
\providecommand{\BIBentryALTinterwordstretchfactor}{4}
\providecommand{\BIBentryALTinterwordspacing}{\spaceskip=\fontdimen2\font plus
\BIBentryALTinterwordstretchfactor\fontdimen3\font minus
  \fontdimen4\font\relax}
\providecommand{\BIBforeignlanguage}[2]{{%
\expandafter\ifx\csname l@#1\endcsname\relax
\typeout{** WARNING: IEEEtran.bst: No hyphenation pattern has been}%
\typeout{** loaded for the language `#1'. Using the pattern for}%
\typeout{** the default language instead.}%
\else
\language=\csname l@#1\endcsname
\fi
#2}}
\providecommand{\BIBdecl}{\relax}
\BIBdecl

\bibitem{wang2017tacotron}
Y.~Wang, R.~Skerry-Ryan, D.~Stanton, Y.~Wu, R.~Weiss, N.~Jaitly \emph{et~al.},
  ``Tacotron: Towards end-to-end speech synthesis,'' 08 2017, pp. 4006--4010.

\bibitem{shen2018tacotron2}
J.~Shen, R.~Pang, R.~Weiss, M.~Schuster, N.~Jaitly, Z.~Yang \emph{et~al.},
  ``{Natural TTS synthesis by conditioning wavenet on mel spectrogram
  predictions},'' in \emph{Proc. ICASSP}, 2018, pp. 4779--4783.

\bibitem{deepvoice2}
A.~Gibiansky, S.~Arik, G.~Diamos, J.~Miller, K.~Peng, W.~Ping \emph{et~al.},
  ``Deep voice 2: Multi-speaker neural text-to-speech,'' in \emph{Proc.
  Advances in Neural Information Processing Systems}, vol.~30, 2017.

\bibitem{deepvoice3}
W.~Ping, K.~Peng, A.~Gibiansky, S.~O. Arik, A.~Kannan, S.~Narang \emph{et~al.},
  ``Deep voice 3: 2000-speaker neural text-to-speech,'' in \emph{Proc. Int.
  Conf. on Learning Representations (ICLR)}, 2018.

\bibitem{ren2019fastspeech}
Y.~Ren, Y.~Ruan, X.~Tan, T.~Qin, S.~Zhao, Z.~Zhao, and T.-Y. Liu, ``Fastspeech:
  Fast, robust and controllable text to speech,'' in \emph{Proc. Advances in
  Neural Information Processing Systems}, 2019, pp. 3165--3174.

\bibitem{ren2021fastspeech2}
Y.~Ren, C.~Hu, X.~Tan, T.~Qin, S.~Zhao, Z.~Zhao, and T.-Y. Liu, ``Fastspeech 2:
  Fast and high-quality end-to-end text to speech,'' in \emph{Proc. Int. Conf.
  on Learning Representations (ICLR)}, 2021.

\bibitem{fastpitch}
A.~{\L}a{\'n}cucki, ``Fastpitch: Parallel text-to-speech with pitch
  prediction,'' \emph{arXiv preprint arXiv:2006.06873}, 2020.

\bibitem{jsbae}
J.~S. Bae, H.~Bae, Y.~S. Joo, J.~Lee, G.~H. Lee, and H.~Y. Cho, ``Speaking
  speed control of end-to-end speech synthesis using sentence-level
  conditioning,'' in \emph{Proc. Interspeech}, 2020, pp. 4402--4406.

\bibitem{oord2016wavenet}
A.~V.~D. Oord, S.~Dieleman, H.~Zen, K.~Simonyan, O.~Vinyals, A.~Graves
  \emph{et~al.}, ``Wavenet: A generative model for raw audio,'' in \emph{arXiv
  preprint arXiv:1609.03499}, 2016.

\bibitem{oord2018parawavenet}
A.~V.~D. Oord, I.~Li, I.~Babuschkin, K.~Simonyan, O.~Vinyals, K.~Kavukcuoglu
  \emph{et~al.}, ``{Parallel WaveNet: Fast high-fidelity speech synthesis},''
  in \emph{Proc. Int. Conf. on Machine Learning (ICML)}, vol.~80, 2018, pp.
  3918--3926.

\bibitem{yang2020vocgan}
J.~Yang, J.~Lee, Y.~Kim, H.-Y. Cho, and I.~Kim, ``{VocGAN: A High-Fidelity
  Real-Time Vocoder with a Hierarchically-Nested Adversarial Network},''
  \emph{Proc. Interspeech}, pp. 200--204, 2020.

\bibitem{yamamoto2020parallel}
R.~Yamamoto, E.~Song, and J.-M. Kim, ``{Parallel WaveGAN: A fast waveform
  generation model based on generative adversarial networks with
  multi-resolution spectrogram},'' in \emph{Proc. ICASSP}, 2020, pp.
  6199--6203.

\bibitem{kumar2019melgan}
K.~Kumar, R.~Kumar, T.~de~Boissiere, L.~Gestin, W.~Z. Teoh, J.~Sotelo
  \emph{et~al.}, ``{MelGAN: Generative adversarial networks for conditional
  waveform synthesis},'' in \emph{Proc. Advances in Neural Information
  Processing Systems}, 2019, pp. 14\,881--14\,892.

\bibitem{hifigan}
J.~Kong, J.~Kim, and J.~Bae, ``{HiFi-GAN: Generative adversarial networks for
  efficient and high fidelity speech synthesis},'' in \emph{Proc. Advances in
  Neural Information Processing Systems}, 2020, pp. 17\,022--17\,033.

\bibitem{libri_tts}
H.~Zen, V.~Dang, R.~Clark, Y.~Zhang, R.~J. Weiss, Y.~Jia \emph{et~al.},
  ``{LibriTTS: A corpus derived from LibriSpeech for text-to-speech},'' in
  \emph{Proc. Interspeech 2019}, 2019, pp. 1526--1530.

\bibitem{multispeech}
M.~Chen, X.~Tan, Y.~Ren, J.~Xu, H.~Sun, S.~Zhao, and T.~Qin, ``{MultiSpeech:
  Multi-speaker text to speech with Transformer},'' in \emph{Proc.
  Interspeech}, 2020, pp. 4024--4028.

\bibitem{towards_univ_tts}
J.~Yang and L.~He, ``{Towards universal text-to-speech},'' in \emph{Proc.
  Interspeech}, 2020, pp. 3171--3175.

\bibitem{adaspeech}
M.~Chen, X.~Tan, B.~Li, Y.~Liu, T.~Qin, sheng zhao, and T.-Y. Liu, ``Adaspeech:
  Adaptive text to speech for custom voice,'' in \emph{Proc. Int. Conf. on
  Learning Representations (ICLR)}, 2021.

\bibitem{semi_sup_improving_data_eff}
Y.~{Chung}, Y.~{Wang}, W.~{Hsu}, Y.~{Zhang}, and R.~J. {Skerry-Ryan},
  ``Semi-supervised training for improving data efficiency in end-to-end speech
  synthesis,'' in \emph{Proc. ICASSP}, 2019, pp. 6940--6944.

\bibitem{chen2018sample_efficient}
Y.~Chen, Y.~Assael, B.~Shillingford, D.~Budden, S.~Reed, H.~Zen \emph{et~al.},
  ``Sample efficient adaptive text-to-speech,'' in \emph{Proc. Int. Conf. on
  Learning Representations (ICLR)}, 2019.

\bibitem{neural_voice_cloning}
S.~Arik, J.~Chen, K.~Peng, W.~Ping, and Y.~Zhou, ``Neural voice cloning with a
  few samples,'' in \emph{Proc. Advances in Neural Information Processing
  Systems}, vol.~31, 2018.

\bibitem{high_quality_lightweight}
Z.~Kons, S.~Shechtman, A.~Sorin, C.~Rabinovitz, and R.~Hoory, ``{High Quality,
  Lightweight and Adaptable TTS Using LPCNet},'' in \emph{Proc. Interspeech},
  2019, pp. 176--180.

\bibitem{boffintts}
H.~B. {Moss}, V.~{Aggarwal}, N.~{Prateek}, J.~{González}, and
  R.~{Barra-Chicote}, ``Boffin tts: Few-shot speaker adaptation by bayesian
  optimization,'' in \emph{Proc. ICASSP}, 2020, pp. 7639--7643.

\bibitem{zhang2020adadurian}
Z.~Zhang, Q.~Tian, H.~Lu, L.-H. Chen, and S.~Liu, ``Adadurian: Few-shot
  adaptation for neural text-to-speech with durian,'' 2020.

\bibitem{transfer_learning_from_ASV}
Y.~Jia, Y.~Zhang, R.~Weiss, Q.~Wang, J.~Shen, F.~Ren \emph{et~al.}, ``Transfer
  learning from speaker verification to multispeaker text-to-speech
  synthesis,'' in \emph{Proc. Advances in Neural Information Processing
  Systems}, vol.~31, 2018.

\bibitem{zero_shot_MS}
E.~{Cooper}, C.~I. {Lai}, Y.~{Yasuda}, F.~{Fang}, X.~{Wang}, N.~{Chen}, and
  J.~{Yamagishi}, ``Zero-shot multi-speaker text-to-speech with
  state-of-the-art neural speaker embeddings,'' in \emph{Proc. ICASSP}, 2020,
  pp. 6184--6188.

\bibitem{song2020speakeradaptive}
E.~Song, J.-S. Kim, K.~Byun, and H.-G. Kang, ``Speaker-adaptive neural vocoders
  for parametric speech synthesis systems,'' in \emph{2020 IEEE 22nd
  International Workshop on Multimedia Signal Processing (MMSP)}.\hskip 1em
  plus 0.5em minus 0.4em\relax IEEE, 2020, pp. 1--5.

\bibitem{wang2018pix2pixhd}
T.-C. Wang, M.-Y. Liu, J.-Y. Zhu, A.~Tao, J.~Kautz \emph{et~al.},
  ``High-resolution image synthesis and semantic manipulation with conditional
  gans,'' in \emph{Proc. of the IEEE conf. on computer vision and pattern
  recognition}, 2018, pp. 8798--8807.

\bibitem{zhang2018stackgan++}
H.~Zhang, T.~Xu, H.~Li, S.~Zhang, X.~Wang, X.~Huang \emph{et~al.},
  ``Stackgan++: Realistic image synthesis with stacked generative adversarial
  networks,'' \emph{IEEE transactions on pattern analysis and machine
  intelligence}, vol.~41, no.~8, pp. 1947--1962, 2018.

\bibitem{salimans2016improved}
T.~Salimans, I.~Goodfellow, W.~Zaremba, V.~Cheung, A.~Radford, and X.~Chen,
  ``{Improved techniques for training GANs},'' in \emph{Proc. Int. Conf. on
  Neural Information Processing Systems}, 2016, pp. 2234--2242.

\bibitem{adv_korean_singing}
J.~Lee, H.-S. Choi, C.-B. Jeon, J.~Koo, and K.~Lee, ``Adversarially trained
  end-to-end korean singing voice synthesis system,'' in \emph{Proc.
  Interspeech}, 2019, pp. 2588--2592.

\bibitem{sheng2019enhancer}
L.~Sheng, D.-Y. Huang, and E.~N. Pavlovskiy, ``High-quality speech synthesis
  using super-resolution mel-spectrogram,'' \emph{arXiv preprint
  arXiv:1912.01167}, 2019.

\bibitem{transformer}
A.~Vaswani, N.~Shazeer, N.~Parmar, J.~Uszkoreit, L.~Jones, A.~N. Gomez
  \emph{et~al.}, ``Attention is all you need,'' in \emph{Proc. Advances in
  neural information processing systems}, 2017, pp. 5998--6008.

\bibitem{mao2017least}
X.~Mao, Q.~Li, H.~Xie, R.~Y. Lau, Z.~Wang, and S.~Paul~Smolley, ``Least squares
  generative adversarial networks,'' in \emph{Proc. of the IEEE Int. Conf. on
  Computer Vision}, 2017, pp. 2794--2802.

\bibitem{kingma2014adam}
D.~P. Kingma and J.~Ba, ``Adam: A method for stochastic optimization,'' in
  \emph{Proc. Int. Conf. on Learning Representations (ICLR)}, 2015.

\bibitem{li2019transformertts}
N.~Li, S.~Liu, Y.~Liu, S.~Zhao, and M.~Liu, ``Neural speech synthesis with
  transformer network,'' in \emph{Proc. of the AAAI Conference on Artificial
  Intelligence}, vol.~33, no.~01, 2019, pp. 6706--6713.

\bibitem{chorowski2015attention}
J.~Chorowski, D.~Bahdanau, D.~Serdyuk, K.~Cho, and Y.~Bengio, ``Attention-based
  models for speech recognition,'' in \emph{NIPS}, 2015.

\bibitem{praat}
P.~Boersma, ``Accurate short-term analysis of the fundamental frequency and the
  harmonics-to-noise ratio of a sampled sound,'' in \emph{Proc. institute of
  phonetic sciences}, 1993, pp. 97--110.

\bibitem{pesq}
A.~W. Rix, J.~G. Beerends, M.~P. Hollier, and A.~P. Hekstra, ``Perceptual
  evaluation of speech quality (pesq)-a new method for speech quality
  assessment of telephone networks and codecs,'' in \emph{Proc. ICASSP},
  vol.~2, 2001, pp. 749--752.

\end{thebibliography}
